\documentclass[twocolumn,showpacs,showkeys]{revtex4}
\usepackage{graphicx}
\usepackage{bm}
\usepackage{color}
\usepackage{amsmath}
\usepackage{natbib}

\begin{document}

\title{Hydrodynamics of the atomic Bose-Einstein condensate  beyond the mean-field approximation: a mini-review}

\author{Pavel A. Andreev}
\email{andreevpa@physics.msu.ru}
\affiliation{Faculty of physics, Lomonosov Moscow State University, Moscow, Russian Federation, 119991.}

\date{\today}

\begin{abstract}
Several hydrodynamic models the atomic Bose-Einstein condensate beyond the mean-field approximation are discussed together from one point of view.
All these models are derived from microscopic quantum description.
The derivation is made within the many-particle quantum hydrodynamics method suggested by L. Kuz'menkov.
The derivation is demonstrated and discussed for the mean-field regime
revealing the Gross-Pitaevskii equation as the simplest illustration.
It appears in the first order by the interaction radius.
Generalization of the hydrodynamic Euler equation obtained in the third order by the interaction radius are discussed.
It includes the contribution of the isotropic short-range interaction
presented by the third space derivative of the square of concentration.
The Euler equation also includes the contribution of the anisotropic part of the short-range interaction
proportional to the second order spherical function.
Systematic account of the quantum fluctuations in terms of the many-particle quantum hydrodynamics method
requires the extension of the set of hydrodynamic equations from the couple continuity and Euler equations
to the set of four equations which also includes the pressure evolution equation
and the evolution equation for the third rank analog of pressure.
The pressure evolution equation contains no interaction contribution in the first order by the interaction radius.
The source of the quantum fluctuations is in the interaction caused term in the third rank tensor evolution equation
which is obtained in the first order by the interaction radius.
\end{abstract}

\pacs{03.75.Hh, 03.75.Kk, 67.85.Pq}
\keywords{quantum hydrodynamics, pressure evolution equation, extended hydrodynamics, quantum fluctuations, mean-field approximation.}


\maketitle


\section{Introduction}

Majority of physical effects in the Bose-Einstein condensate (BEC) of ultracold atomic gases
can be described by the Gross-Pitaevskii (GP) equation \cite{Dalfovo RMP 99}, \cite{Szirmai PRA 12}, \cite{Stamper-Kurn RMP 13}.
Which is the non-linear Schrodinger equation with the cubic nonlinearity
found in the mean-field approximation for the short-range interaction (SRI).
Here, the interaction is reduced to the scattering of atoms with the zeroth transfer of the momentum.
So, the interaction is characterized by the s-wave scattering length $a$.
The GP equation can be represented in the form of hydrodynamic equations for the curl-free "fluid".

In past years, there is the growth of the number of phenomena in neutral BECs which require models obtained beyond the mean field approximation.
Therefore, few generalized approaches are presented in literature.

The famous example of the analytical model obtained beyond the mean-field approximation is the model of quantum fluctuations in BECs \cite{Lima PRA 12}.
It includes the presence of bosons in the excited states at the zero temperature \cite{Lima PRA 12}.
The generalization of this model for the dipolar BEC (see \cite{Lahaye RPP 09} and \cite{Lahaye Nat 07})
is the crucial instrument in the modeling of the quantum-droplets formation
\cite{Baillie PRA 16}, \cite{Ferrier-Barbut PRL 16}, \cite{Kadau Pfau Nature 16},
which were observed experimentally.
The contribution of quantum fluctuations can be approximately presented in the generalized GP equation,
which contains the fourth-degree nonlinearity along with the third degree nonlinearity.

Characteristic feature of the interaction of neutral atoms is the short range of the interaction (except the dipolar BEC).
This short range is comparable with the radius of atoms,
but it is small in compare with the average interparticle distance
or the characteristic length of the collective effects
such as the wavelength of waves and the width of solitons.
Therefore, we can expand general functions describing interaction on the small parameter
which is the nontrivial interparticle distance,
where the interaction is nonzero.
At the larger distances the potential goes to zero,
so they are not relevant.
The main contribution appears in the first order on the small parameter (the first order by the interaction radius).
It leads to the GP equation.

However, we can consider the next orders of the expansion.
The non-zero contribution appears in the third order by the interaction radius \cite{Andreev PRA08}.
It is obtained under assumption
that all bosons are in the lowest energy state.
Therefore,
this generalization of the GP model does not related to the quantum fluctuations described above.
The terms existing in the third order by the interaction radius (TOIR) are easily presented in terms of hydrodynamic model,
but this approximation does not allow to obtain any non-linear Schrodinger equation.
The TOIR approximation is presented by the higher derivatives of the concentration in the Euler equation.
Moreover, it introduces the second interaction constant.
The interaction constant in the GP equation is the zeroth moment of the potential of interatomic interaction
\begin{equation} \label{BECBMFMR2020 def g} g=\int d\textbf{r}U(r). \end{equation}
The TOIR approximation gives the second interaction constant
which is proportional to the second moment of the potential
\begin{equation} \label{BECBMFMR2020 g 2 TOIR prop} g_{2}\sim\int d\textbf{r}r^{2}U(r). \end{equation}
The notion of moment of the function is similar to the similar notion in the probability theory,
but the similarity just in the structure of the integrals.
Different approach to the nonlocal generalization of the GP equation is presented in Refs. \cite{Braaten PRA 01}, \cite{Rosanov PLA 02}

The term in the Euler equation appearing in the TOIR approximation leads to the novel type of solitons in BECs
\cite{Andreev MPL B 12}, \cite{Andreev LP 19}.
Actually, it creates conditions for existence of the bright soliton in the repulsive BECs \cite{Andreev MPL B 12}.
Similar phenomenon is found experimentally in Ref. \cite{Wang NJP 14}.

If we include the anisotropy of the SRI
it leaves
no trace in the mean-field model obtained in the first order by the interaction radius (the GP approximation).
However, the part of anisotropy described by the spherical function of the second order $Y_{20}$
leads to the additional term in the TOIR approximation \cite{Andreev LP 19}.
It also introduces the third interaction constant.

It is noticed that the collisions of solitons in BECs requires modeling of BEC beyond mean-field approximation \cite{Katsimiga NJP 17 01}.
To this end, corresponding models are developed \cite{Katsimiga NJP 17 02}, \cite{Katsimiga PRA 18}, \cite{Mistakidis NJP 18}.
They are based on the microscopic dynamics traced through the numerical methods.

It is possible to derive the quantum fluctuations purely from the quantum hydrodynamics based on the microscopic quantum dynamics of bosons.
It requires to extend the set of hydrodynamic equations from two equations (the continuity and Euler equations)
to the set of four equations (the continuity equation, Euler equation, the pressure evolution equation, and the third rank tensor evolution equation).
The pressure is the second rank tensor constructed as the average of the product of two velocities in the comoving frame.
The third rank tensor is the flux of pressure in the comoving frame,
hence it is the average of the product of three velocities.
The nonzero velocities in the comoving frame is the trace of the particles in the excited states.
The third rank tensor evolution equation contains the contribution of the SRI
which has nonzero value in the first order by the interaction radius at the zero temperature.
It is proportional to novel (fourth) interaction constant,
which is proportional to the zeroth moment of the second derivative of the potential
\begin{equation} \label{BECBMFMR2020 def g 4 FOIR Q} g_{4}\sim\int d\textbf{r}U''(r). \end{equation}
The third rank tensor evolution equation includes the source existing at the zero temperature caused by the SRI.
Therefore, it is interpreted as the quantum fluctuations.

We described some results obtained within the quantum hydrodynamics method.
However, the "generalized hydrodynamics" is also method of description of collective phenomena including quantum effects
\cite{Ruggiero arxiv 19}, \cite{Bertini PRL 16}.
Particularly, the quantum fluctuations in BECs are discussed within this method \cite{Ruggiero arxiv 19}.

Mentioned hydrodynamic models are described in more details in the text below.

This paper is organized as follows.
In Sec. II describes the mean-field approximation including the derivation of the GP equation by the quantum hydrodynamic method.
In Sec. III shows the generalized equations of quantum hydrodynamics obtained beyond mean-field approximation.
In Sec. IV a brief summary of obtained results is presented.

\section{Mean-field model of BEC}

The paper is focused on the models of atomic BEC beyond mean-field approximation.
However, we start of paper with description of the mean-field approximation as the solid background for the generalizations.

The BEC is the collection of bosons being in the single quantum state with the minimal energy.

Weakly interacting atomic BEC of neutral particles interacting via the SRI can be approximately described
by the well-known GP equation \cite{Dalfovo RMP 99}:
\begin{equation} \label{BECBMFMR2020 GP}
\imath\hbar\partial_{t}\Phi=-\frac{\hbar^{2}}{2m}\triangle\Phi+V_{ext}\Phi +g\mid\Phi\mid^{2}\Phi, \end{equation}
where
$\hbar$ is the Planck constant,
$m$ is the mass of particle,
$\imath$ is the imaginary unit,
$\partial_{t}$ is the derivative on time $t$,
$\triangle$ is the Laplace operator,
$V_{ext}$ is the potential of the external field,
$g$ is the interaction constant (\ref{BECBMFMR2020 def g}), which can be interpreted via the scattering length $a$: $g=4\pi\hbar^{2}a/m$,
$\Psi$ is the effective macroscopic wave function.
It is the non-linear Schrodinger equation,
where the nonlinearity is caused by the interaction of bosons.
Here, the interaction is presented approximately via the major term.
GP equation (\ref{BECBMFMR2020 GP}) can be represented in the form of hydrodynamic equation.

Well-known derivation of the GP equation
from the Schrodinger equation written via psi-operators,
which corresponds to the quantum mechanics in the representation of the second quantization \cite{Landau IX}.
In this derivation the psi-operators are reduced to the functions of coordinate and time
and the interaction potential is replaced by the quasi-potential,
to obtain the description of the BEC.

Another derivation of the GP equation directly from the many-particle Schrodinger equation in the coordinate representation
is made via the quantum hydrodynamics method \cite{Andreev PRA08}.
Let us present this derivation here.

\subsection{Derivation of the GP equation within the quantum hydrodynamic method}

The derivation of the GP equation is splitted on two parts.
First, derivation of general form of hydrodynamic equations is presented.
Second, the GP equation is found at the approximate analysis of term describing the interaction.

\subsubsection{Derivation of the continuity and Euler equation in general form from the microscopic quantum motion}

Microscopic quantum dynamics of system of particles is described within the Schrodinger equation $\imath\hbar\partial_{t}\psi(R,t)=\hat{H}\psi(R,t)$.
Different quantum systems differ from each other by the explicit form of the Hamiltonian $\hat{H}$
and by the symmetry of wave function $\psi(R,t)$ relatively to permutations of arguments.

In this paper, we consider single species neutral spin-0 bosons.
Therefore, the wave function is symmetric relatively all permutations of arguments.
Masses of all particles are equal to each other $m_{i}\equiv m$.
System of interacting bosons placed in the external field with potential $V_{ext}(\textbf{r}_{i},t)$ is described by the following Hamiltonian
\begin{equation}\label{BECBMFMR2020 microscopic Hamiltonian}
\hat{H}=\sum_{i=1}^{N}\frac{\hat{\textbf{p}}^{2}_{i}}{2m_{i}}+\sum_{i=1}^{N}V_{ext}(\textbf{r}_{i},t)
+\frac{1}{2}\sum_{i,j,i\neq j}U(\mid \textbf{r}_{i}-\textbf{r}_{j}\mid) ,\end{equation}
where $N$ is the number of particles,
$\hat{\textbf{p}}_{i}=-\imath\hbar\nabla_{i}$ is the momentum operator of the i-th particle,
and $U_{ij}=U(\mid\textbf{r}_{i}-\textbf{r}_{j}\mid)$ is the interaction potential.
The number of particles is included in the wave function via the number of arguments
$R=\{\textbf{r}_{1}, ..., \textbf{r}_{N}\}$.

We present the derivation in accordance with the many-particle quantum hydrodynamics method suggested by L. Kuz'menkov \cite{Maksimov QHM 99-01}
(see also Ref. \cite{Andreev PRB 11}),
which is adopted for the ultracold gases in Refs. \cite{Andreev PRA08}, \cite{Andreev IJMPB 13}.

The concentration (number density) of particles is the simplest fundamental hydrodynamic function for the classical and quantum description.
It can be defined in the vicinity of the point $\textbf{r}$ of the physical space as the operator
$\hat{n}=\sum_{i=1}^{N}\delta(\textbf{r}-\textbf{r}_{i})$ averaged with the many-particle microscopic wave function $\psi(R,t)$:
\begin{equation}\label{BECBMFMR2020 def concentration}
n(\textbf{r},t)=\int dR\sum_{i}\delta(\textbf{r}-\textbf{r}_{i})\psi^{*}(R,t)\psi(R,t),\end{equation}
where $dR=\prod_{i=1}^{N}d\textbf{r}_{i}$ is the element of volume in $3N$ dimensional configurational space.

Operator $\hat{n}$ is obtained by quantization of the classic microscopic concentration of particles,
which is the sum of delta-functions.
The coordinates of classic particles are replaced by the operator of coordinate in accordance with the quantization rule.
The coordinate operator of $i$-th particle has trivial form in the coordinate representation of quantum mechanics
$\hat{\textbf{r}}_{i}=\textbf{r}_{i}$.

Let us point out that the delta-function is the traditional element of quantum description.
Major quantum example is the eigen-function of the coordinate operator in the coordinate representation of quantum mechanics.

Hydrodynamics describe the time evolution of the physical systems.
Therefore, it is necessary to derive equation for the concentration evolution.
Number of particles is conserves in our description
since we neglect the molecular formation
which happens in the traps and leads to destruction of the condensate cloud.
Hence, it is expected that the concentration satisfies the continuity equation.
However, it is essential to derive the continuity equation
since the derivation gives us the explicit form of the second hydrodynamic function in terms of the many-particle wave function $\psi(R,t)$.

Consider the time derivative of the concentration (\ref{BECBMFMR2020 def concentration}).
The time derivative acts on the argument of wave functions
which can be taken from the Schrodinger equation with Hamiltonian (\ref{BECBMFMR2020 microscopic Hamiltonian}).
After straightforward calculations
we obtain that
the term under consideration is the divergence of the following vector function
$$j^{\alpha}(\textbf{r},t)=\int dR\sum_{i}\delta(\textbf{r}-\textbf{r}_{i})\times$$
\begin{equation}\label{BECBMFMR2020 def current}
\times\frac{1}{2m_{i}}[(\hat{p}^{\alpha}_{i}\psi)^{*}(R,t)\psi(R,t) +\psi^{*}(R,t)(\hat{p}^{\alpha}_{i}\psi)(R,t)],\end{equation}
which is the particle current proportional to the local momentum density of the bosons.
We find that the concentration of bosons (\ref{BECBMFMR2020 def concentration}) satisfies the continuity equation
\begin{equation}\label{BECBMFMR2020 continuity via j} \partial_{t}n+\nabla \textbf{j}=0 \end{equation}
with the current (\ref{BECBMFMR2020 def current}).

At this step we have the single unknown hydrodynamic function it is the current of particles (\ref{BECBMFMR2020 def current}).
Equation for its evolution can be found using the Schrodinger equation with Hamiltonian (\ref{BECBMFMR2020 microscopic Hamiltonian}).
Let us present the result of calculation
$$\partial_{t}j^{\alpha}(\textbf{r},t)+\frac{1}{m}\partial_{\beta}\Pi^{\alpha\beta}(\textbf{r},t)
=-\frac{1}{m}n(\textbf{r},t)\nabla^{\alpha}V_{ext}(\textbf{r},t)$$
\begin{equation}\label{BECBMFMR2020 momentum balance evol eq}
-\frac{1}{m}\int d\textbf{r}'(\nabla^{\alpha}U(\textbf{r},\textbf{r}'))n_{2}(\textbf{r},\textbf{r}',t).
\end{equation}

All hydrodynamic functions are defined as the fields in the three-dimensional physical space changing with time.
Their arguments is the same
it is $(\textbf{r},t)$.
Therefore, we mostly drop the arguments of these functions.
Correlation functions like $n_{2}(\textbf{r},\textbf{r}',t)$ has more complex structure of arguments,
hence we keep them explicit.

The left-hand side of equation (\ref{BECBMFMR2020 momentum balance evol eq}) contains the kinetic function
while the right-hand side contains the interaction.
We keep this structure for all hydrodynamic equations below.
The first term on the right-hand side describes the interaction of the bosons with the external field.
The second term presents the interaction between bosons.

Tensor $\Pi^{\alpha\beta}$ is the quantum tensor of the momentum flux:
$$\Pi^{\alpha\beta}(\textbf{r},t)=
\int dR\sum_{i}\delta(\textbf{r}-\textbf{r}_{i})\frac{1}{4m_{i}}[\psi^{*}(R,t)(\hat{p}^{\alpha}_{i}\hat{p}^{\beta}_{i}\psi)(R,t)$$
\begin{equation} \label{BECBMFMR2020 def Pi}
+(\hat{p}^{\alpha}_{i}\psi)^{*}(R,t)(\hat{p}^{\beta}_{i}\psi)(R,t)+c.c.], \end{equation}
where $c.c.$ is the complex conjugation.
It is the symmetric tensor $\Pi^{\alpha\beta}=\Pi^{\beta\alpha}$.

The interaction between the particles in equation (\ref{BECBMFMR2020 momentum balance evol eq})
is expressed via the two-particle concentration
$n_{2}(\textbf{r},\textbf{r}',t)$ normalized over $N(N-1)$:
\begin{equation} \label{BECBMFMR2020 def n2}
n_{2}(\textbf{r},\textbf{r}',t)=\int dR\sum_{i,j,i\neq j}\delta(\textbf{r}-\textbf{r}_{i})\delta(\textbf{r}'-\textbf{r}_{j})\psi^{*}(R,t)\psi(R,t).\end{equation}
Function $n_{2}(\textbf{r},\textbf{r}',t)$ splits on the product of concentrations $n(\textbf{r},t)\cdot n(\textbf{r}',t)$
if there is no two-particle correlations.

Equations (\ref{BECBMFMR2020 continuity via j}) and (\ref{BECBMFMR2020 momentum balance evol eq})
are the fundamental pair of hydrodynamic equations.
They obtained for bosons with two-particle potential interaction,
but same equations can be obtained for the fermions.
Strength of interaction or the radius of the influence of the potential are not included in the derivation.

Our goal is the derivation of the GP equation for the weakly interacting bosons with the SRI between bosons.
Hence, we need to include these features in the momentum balance equation (\ref{BECBMFMR2020 momentum balance evol eq}).

First we include the short range nature of the boson-boson interaction.

Interaction in equation (\ref{BECBMFMR2020 momentum balance evol eq}) is written via the two-particle concentration (\ref{BECBMFMR2020 def n2}).
To consider the small area of influence of the potential
we need to rewrite the force field via the microscopic wave function $\psi(R,t)$.
It helps to trace the symmetry relatively permutations of particles.
Original form of the force field is
\begin{equation} \label{BECBMFMR2020 FF1} F^{\alpha}=-\int dR\sum_{i,j.i\neq j} \delta(\textbf{r}-\textbf{r}_{i})(\nabla^{\alpha}_{i}U(\textbf{r}_{ij}))\psi^{*}(R,t)\psi(R,t).\end{equation}
This expression does not depend on $i$ and $j$.
Replacing indexes $i$ and $j$ we find different form of the force field
\begin{equation} \label{BECBMFMR2020 FF2} F^{\alpha}=\int dR\sum_{i,j.i\neq j} \delta(\textbf{r}-\textbf{r}_{j}))(\nabla^{\alpha}_{i}U(\textbf{r}_{ij}))\psi^{*}(R,t)\psi(R,t).\end{equation}
To include the symmetries we consider the half sum of these expressions
$$F^{\alpha}=-\frac{1}{2}\int dR\sum_{i,j.i\neq j} (\delta(\textbf{r}-\textbf{r}_{i})-\delta(\textbf{r}-\textbf{r}_{j}))\times$$
\begin{equation} \label{BECBMFMR2020 FF3 symm} \times(\nabla^{\alpha}_{i}U(\textbf{r}_{ij}))\psi^{*}(R,t)\psi(R,t).\end{equation}

Small rance of the potential action $U(\textbf{r}_{ij})$ corresponds to nontrivial value of potential $U(\textbf{r}_{ij})$ at small $\textbf{r}_{ij}$.
The SRI potential goes to zero at the large interparticle distance $\textbf{r}_{ij}$.
We need to trace areas of small and large interparticle distances $\textbf{r}_{ij}$
in all functions under the integral in the force field expression (\ref{BECBMFMR2020 FF3 symm}).
Therefore, we introduce the following variables:
the interparticle distance for two particles $\textbf{r}_{ij}$,
and the center of mass for the pair of particles $\textbf{R}_{ij}$,
which are defined as follows
\begin{equation} \label{BECBMFMR2020 def r R}
\begin{array}{ccc}
\textbf{R}_{ij}=\frac{1}{2}(\textbf{r}_{i}+\textbf{r}_{j}) ,
& \textbf{r}_{ij}=\textbf{r}_{i}-\textbf{r}_{j}
\end{array}.
\end{equation}

Replacing $\textbf{r}_{i}$ and $\textbf{r}_{j}$
via the relative distance $\textbf{r}_{ij}$,
and the center of mass $\textbf{R}_{ij}$.
It includes the arguments of the wave function
$\psi(R,t)=\psi(\textbf{r}_{1},..,\textbf{r}_{i},..,\textbf{r}_{j},...,\textbf{r}_{N},t)$.

Nonzero value of the function under integral in (\ref{BECBMFMR2020 FF3 symm}) exists at the small relative distance $\textbf{r}_{ij}$.
Therefore, we can expand the function under the integral on relative distance $\textbf{r}_{ij}$ in the Taylor series.
The zeroth term of the expansion is equal to zero.
Main contribution is given by the first term of expansion.
The result can be written in the following form
\begin{equation} \label{BECBMFMR2020 F via sigma} F^{\alpha}=-\partial_{\beta}\sigma^{\alpha\beta} \end{equation}
where
$$\sigma^{\alpha\beta}(\textbf{r},t)=-\frac{1}{2}\int dR\sum_{i,j.i\neq j}\delta(\textbf{r}-\textbf{R}_{ij})\times$$
\begin{equation} \label{BECBMFMR2020 sigma in 1 order} \times\frac{r^{\alpha}_{ij}r^{\beta}_{ij}}{\mid\textbf{r}_{ij}\mid}\frac{\partial U(\textbf{r}_{ij})}{\partial\mid\textbf{r}_{ij}\mid}\psi^{*}(R,t)\psi(R,t)\end{equation}
is the quantum stress tensor.

The explicit form of the quantum stress tensor (\ref{BECBMFMR2020 sigma in 1 order}) shows that
it is symmetric with respect to permutation of indexes $\alpha,\beta$:
$\sigma^{\alpha\beta}=\sigma^{\beta\alpha}$.

The momentum balance equation (\ref{BECBMFMR2020 momentum balance evol eq}) can be rewritten via the quantum stress tensor
\begin{equation}\label{BECBMFMR2020 momentum balance evol eq via sigma}\partial_{t}j^{\alpha}+\frac{1}{m}\partial_{\beta}
(\Pi^{\alpha\beta}+\sigma^{\alpha\beta})=-\frac{1}{m}n\nabla^{\alpha}V_{ext},\end{equation}
where the interparticle interaction is placed on the left-hand side.

Tensor $\Pi^{\alpha\beta}$ is the kinetic momentum flux.
It can be pictured in the following way.
Consider some area of space surrounded with some surface.
Motion of particles leads to the loss and acquisition of particles
which cross the surface during their motion.
Each particle carries the momentum.
Hence, the loss (acquisition) of particles leads to the loss (acquisition) of the momentum located in the considered area of space.
The change of momentum is not related to the interaction.
Same mechanism exists in the ideal gas.
The quantum stress tensor $\sigma^{\alpha\beta}$ can be interpreted as the momentum flux caused by the interaction.
The interaction of particles within the considered area of space does not change the full momentum in this area.
However, the interaction of particles being inside with particles being outside gives the change of momentum in this area of space.
We consider the SRI between particles.
Hence, the particles located near the surface can interact and transfer the momentum across the surface.
The superposition $\Pi^{\alpha\beta}+\sigma^{\alpha\beta}$ gives the full momentum flux.

Expression (\ref{BECBMFMR2020 sigma in 1 order}) is found for the arbitrary strength of interaction.
However, tensor $\sigma^{\alpha\beta}$ in form (\ref{BECBMFMR2020 sigma in 1 order}) has no expression via the hydrodynamic functions.
The application of the weak strength of interaction leads
to the representation of the quantum stress tensor $\sigma^{\alpha\beta}$ via the hydrodynamic functions.
The weakly interacting limit of bosons is presented in the next subsection.
Here, we focus on another item.

The momentum balance equation (\ref{BECBMFMR2020 momentum balance evol eq})
or (\ref{BECBMFMR2020 momentum balance evol eq via sigma}) has small resemblance to the traditional Euler equation.
The Euler equation is written via the hydrodynamic velocity field $\textbf{v}$.
Consequently, we need to make transition of the form of hydrodynamic equation based on the velocity field.

The microscopic many-particle wave function $\psi(R,t)$ is the complex function of the real arguments.
It can be rewritten via two real functions
\begin{equation} \label{BECBMFMR2020 psi exp form} \psi(R,t)=a(R,t) \exp\biggl(\frac{\imath S(R,t)}{\hbar}\biggr),\end{equation}
where
$a(R,t)$ is the amplitude of the wave function,
and
$S(R,t)$ is the phase of the wave function.

Analysis of current (\ref{BECBMFMR2020 def current}) leads to the following structure
\begin{equation} \textbf{v}_{i}(R,t)=\frac{1}{m_{i}}\nabla_{i}S(R,t), \end{equation}
which can be interpreted as the velocity of $i$-th quantum particle.

Velocity field $ \textbf{v}(\textbf{r},t) $ is determined via the current and concentration:
\begin{equation}\label{curr}\textbf{j}=n\textbf{v}.\end{equation}

There is nonzero difference between two introduced function $\textbf{v}_{i}(R,t)$ and $\textbf{v}(\textbf{r},t)$.
It is the velocity of $i$-th quantum particle in the comoving frame
$\textbf{u}_{i}(\textbf{r},R,t)=\textbf{v}_{i}(R,t)-\textbf{v}(\textbf{r},t)$.
It can be interpreted as the quantum analog of the velocity of thermal motion.

Next, we need to separate the thermal motion of the particles with
velocities $\textbf{u}_{i}$ and the motion with the hydrodynamic velocity
$\textbf{v}(\textbf{r},t)$ in the continuity equation and the momentum balance
equation (\ref{BECBMFMR2020 momentum balance evol eq via sigma}).
We obtain to the following equations:
\begin{equation}\label{BECBMFMR2020 continuity eq via v1}
\partial_{t}n+\nabla(n\textbf{v})=0 ,\end{equation}
and
\begin{equation}\label{BECBMFMR2020 Eiler eq v1}
mn(\partial_{t}+\textbf{v}\nabla)v^{\alpha}+\partial_{\beta}(p^{\alpha\beta} +\sigma^{\alpha\beta}+T^{\alpha\beta})
=-n\nabla^{\alpha}V_{ext}.\end{equation}
The introduction of the velocity field in the Euler equation requires some straightforward calculations.

The kinetic momentum flux tensor $\Pi^{\alpha\beta}$ splits on three part during the introduction of the velocity field
\begin{equation}\label{BECBMFMR2020 Pi via n v p T} \Pi^{\alpha\beta}=nv^{\alpha}v^{\beta}+p^{\alpha\beta}+T^{\alpha\beta}. \end{equation}

The second term in expression (\ref{BECBMFMR2020 Pi via n v p T}) is the kinetic pressure
\begin{equation}\label{BECBMFMR2020 def pressure} p^{\alpha\beta}(\textbf{r},t)=\int dR\sum_{i=1}^{N}\delta(\textbf{r}-\textbf{r}_{i})a^{2}(R,t)m_{i}u^{\alpha}_{i}u^{\beta}_{i}.\end{equation}
It is related to the distribution of particles over quantum states with different values of the momentum.

The last term in equation (\ref{BECBMFMR2020 Pi via n v p T}) is the quantum Bohm potential
\begin{equation}\label{BECBMFMR2020 Bom1 def} T^{\alpha\beta}(\textbf{r},t)=-\frac{\hbar^{2}}{2m}\int dR\sum_{i=1}^{N}\delta(\textbf{r}-\textbf{r}_{i})a^{2}(R,t)\frac{\partial^{2}\ln a}{\partial x_{\alpha i}\partial x_{\beta i}}.
\end{equation}
It is proportional to $\hbar^{2}$ and has purely quantum origin.

The quantum Bohm potential can be simplified for the system of noninteracting particles located in one quantum states:
\begin{equation} \label{BECBMFMR2020 Bom2 sigle state}
T^{\alpha\beta}=
-\frac{\hbar^{2}}{4m}\biggl(\partial^{\alpha}\partial^{\beta}n-\frac{1}{n}(\partial^{\alpha}n)(\partial^{\beta}n)\biggr).\end{equation}

Divergence of tensor (\ref{BECBMFMR2020 Bom2 sigle state}) is
usually presented in the form:
\begin{equation} \label{BECBMFMR2020 Bom3 div of T}
\partial_{\beta}T^{\alpha\beta}= -\frac{\hbar^{2}}{2m}n\partial^{\alpha}\frac{\triangle\sqrt{n}}{\sqrt{n}}.
\end{equation}

There is no distribution of particles on different quantum states in the BECs.
Therefore, the pressure is equal to zero
\begin{equation} \label{BECBMFMR2020 p=0}p^{\alpha\beta}=0.\end{equation}
Let us point out that it is the kinetic pressure.
Sometimes, at phenomenological analysis, the pressure is introduced at the superposition of the kinetic pressure $p^{\alpha\beta}$
and the quantum stress tensor $\sigma^{\alpha\beta}$.
Hence, the full pressure $P^{\alpha\beta}=p^{\alpha\beta}+\sigma^{\alpha\beta}$ is nonzero,
but its nonzero part is caused by the interaction located in the quantum stress tensor $\sigma^{\alpha\beta}$.

\subsection{Derivation of the Gross-Pitaevskii equation from the general hydrodynamic equation in the first order by the interaction radius}

For the derivation of the GP equation we need to calculate
the quantum stress tensor (\ref{BECBMFMR2020 sigma in 1 order}) in the weakly interacting regime.

The integral over the interparticle distance in the expression (\ref{BECBMFMR2020 sigma in 1 order})
can be separated from over integrals.
Hence, the quantum stress tensor (\ref{BECBMFMR2020 sigma in 1 order}) can be rewritten in the following form
\begin{equation} \label{BECBMFMR2020 sigma via Tr n2}
\sigma^{\alpha\beta}(\textbf{r},t)=-\frac{1}{2}Tr(n_{2}(\textbf{r},\textbf{r}',t))
\int d\textbf{r}\frac{r^{\alpha}r^{\beta}}{r}\frac{\partial U(r)}{\partial r} ,\end{equation}
where the trace of function of two arguments is used
\begin{equation} \label{BECBMFMR2020} Tr f(\textbf{r},\textbf{r}')=f(\textbf{r},\textbf{r}).\end{equation}

The two-particle concentration (\ref{BECBMFMR2020 def n2}) is rewritten in the following form
$$n_2(\textbf{r},\textbf{r}',t)=N(N-1)
\int dR_{N-2}\langle n_1,n_2,\ldots |\textbf{r},\textbf{r}',R_{N-2},t\rangle  \times$$
\begin{equation}\label{BECBMFMR2020 n2 in repr of occupation numbers}
\times \langle\textbf{r},\textbf{r}',R_{N-2},t |n_1,n_2,\ldots\rangle ,
\end{equation}
where $dR_{N-2}=\displaystyle\prod\limits_{k=3}^{N}d\textbf{r}_k$.

Equation (\ref{BECBMFMR2020 n2 in repr of occupation numbers}) contains the microscopic many-particle wave function $\psi(R,t)$
which is written in the representation of occupation numbers $n_{a}$
(the number of particles in the quantum states described by the set of quantum numbers).
The delta functions in definition (\ref{BECBMFMR2020 def n2}) give projection of coordinates $r_{i}$ and $r_{j}$
(after replacement via relative distance and the expansion for the small relative distances)
in the three-dimension physical space arithmetization with coordinate $\textbf{r}$.
The wave function is written in the following form
$\psi(R,t)=\psi(\textbf{r}_{i},\textbf{r}_{j},R_{N-2},t)
\rightarrow
\psi(\textbf{R}_{ij},\textbf{R}_{ij},R_{N-2},t)
\rightarrow
\langle\textbf{r},\textbf{r}',R_{N-2},t |n_1,n_2\ldots\rangle$

We can present the many-particle wave function as the symmetrized product of single-particle wave functions
for the weakly interacting bosons.
We mark out one single particle wave function
$\: \langle\textbf{r},t | f\rangle \:$
$$ \langle \textbf{r}, \textbf{r}', R_{N-2},t |n_1, n_2 \ldots\rangle$$
\begin{equation}\label{BECBMFMR2020 Expansion one fuhnction}=\sum_f \sqrt{\frac{n_f}{N}} \: \langle\textbf{r},t | f\rangle \:
\langle \textbf{r}', R_{N-2},t |n_1, \ldots (n_f-1),\ldots \rangle,\end{equation}
where $\langle \textbf{r},t|f\rangle =\varphi_f(\textbf{r},t)$
are the single-particle wave functions.

Next, we mark out two single particle wave function s
$$ \langle \textbf{r}, \textbf{r}', R_{N-2},t |n_1, n_2 \ldots\rangle$$
$$=\sum_f \sum_{f', {f'\ne f}}
\sqrt{\frac{n_f}{N}} \sqrt{\frac{n_{f'}}{N-1}} \:
\langle\textbf{r},t | f\rangle \: \langle\textbf{r}',t | f'\rangle \times$$
$$\times\langle R_{N-2},t |n_1, \ldots (n_{f'}-1),\ldots (n_f-1), \ldots \rangle$$
$$+\sum_f \sqrt{\frac{n_f(n_f-1)}{N(N-1)}} \:
\langle\textbf{r},t | f\rangle \times$$
\begin{equation}\label{BECBMFMR2020 Expansion two fuhnctions}
\times\: \langle\textbf{r}',t | f\rangle
\: \langle R_{N-2},t |n_1, \ldots (n_f-2), \ldots \rangle.
\end{equation}
The first term on the right-hand side is for two particles in two different quantum states,
and the second term on the right-hand side is for two particles in the single quantum state.
We need last term only for the BECs.
Moreover, we need one term in the sum existing in the second term.

The expanded wave function (\ref{BECBMFMR2020 Expansion two fuhnctions}) is substituted
in equation (\ref{BECBMFMR2020 n2 in repr of occupation numbers}).
It is necessary to get rid of the residue of $N-2$ particle wave function.
It can be made via the orthogonality conditions:
$$\langle  \ldots (n_{f'}-1), \ldots (n_f-1), \ldots |
\ldots (n_{q'}-1),\ldots (n_q-1),\ldots \rangle=$$
\begin{equation}\label{BECBMFMR2020} =\delta (f-q) \delta (f'-q')
+\delta(f-q')\delta (f'-q)
\end{equation}
for two particles in two different quantum states,
and
\begin{equation}\label{BECBMFMR2020}
\langle n_1, \ldots (n_f-2),\ldots |
  n_1, \ldots (n_q-2),\ldots \rangle=\delta (f-q)
\end{equation}
for two particles in the single quantum state.

Described calculation gives the expression of the two-particle concentration
$$n_2(\textbf{r},\textbf{r}',t)=n(\textbf{r},t)n(\textbf{r}',t)$$
\begin{equation}\label{BECBMFMR2020 n2 via n rho}
+|\rho(\textbf{r},\textbf{r}',t)|^{2}+\sum_{g}n_{g}(n_{g}-1)|\varphi_{g}(\textbf{r},t)|^{2}|\varphi_{g}(\textbf{r}',t)|^{2},
\end{equation}
which is partially presented via the concentration written via the single particle wave function
in accordance with definition (\ref{BECBMFMR2020 def concentration})
and
expansion (\ref{BECBMFMR2020 Expansion one fuhnction})
\begin{equation}\label{BECBMFMR2020 n via varphi}
n(\textbf{r},t)=\sum_{g}n_{g}\varphi_{g}^{*}(\textbf{r},t)\varphi_{g}(\textbf{r},t). \end{equation}
However, equation (\ref{BECBMFMR2020 n2 via n rho}) contains the density matrix written via the single particle wave functions
\begin{equation}\label{BECBMFMR2020 rho via varphi} \rho(\textbf{r},\textbf{r}',t)=\sum_{g}n_{g}\varphi_{g}^{*}(\textbf{r},t)\varphi_{g}(\textbf{r}',t).\end{equation}

The first and second terms in two-particle concentration (\ref{BECBMFMR2020 n2 via n rho}) appear
from the first term in expansion (\ref{BECBMFMR2020 Expansion two fuhnctions}).
They are related to the interaction of two particles being in different quantum states.
It described the interaction of BEc with the normal fluid presented by the bosons in the excited states and the selfaction of the normal fluid.

The last term in equation (\ref{BECBMFMR2020 n2 via n rho}) appear from the last term
in expansion (\ref{BECBMFMR2020 Expansion two fuhnctions}).
It describes the interaction of two particles being in the same quantum state.
Ii describes the selfaction of the normal fluid if two bosons are in the excited states.

The last term in equation (\ref{BECBMFMR2020 n2 via n rho}) describes the selfaction of BEC if two bosons are in the lowest energy quantum state.

Next, we use equation (\ref{BECBMFMR2020 n2 via n rho}) for calculation of the quantum stress tensor $\sigma^{\alpha\beta}$
given by equation (\ref{BECBMFMR2020 sigma via Tr n2}).
For this step we take trace of two-particle concentration (\ref{BECBMFMR2020 n2 via n rho})
and find the expression for the quantum stress tensor
\begin{equation}\label{BECBMFMR2020 sigma via n BEC and nf} \sigma^{\alpha\beta}(\textbf{r},t)=
\frac{1}{2}g\delta^{\alpha\beta}\biggl[2n^{2}(\textbf{r},t)
+\sum_{g}n_{g}(n_{g}-1)|\varphi_{g}(\textbf{r},t)|^{4}\biggr],\end{equation}
where we use that
the trace of density matrix is equal to the concentration.
The integral over the relative distance demonstrated in equation (\ref{BECBMFMR2020 sigma via Tr n2}) is represented
via the interaction constant $g\delta^{\alpha\beta}$
which appears in the following form
\begin{equation}\label{BECBMFMR2020 def g original} g=-\frac{4\pi}{3}\int
dr(r)^{3}\frac{\partial U(r)}{\partial r}. \end{equation}
We obtain traditional form of the interaction constant (\ref{BECBMFMR2020 def g}) by integration by parts,
where we use that
$r^{3}U(r)$ tends to zero at $r$ tending to zero and to infinity.

Let us discuss equation (\ref{BECBMFMR2020 sigma via n BEC and nf}) for the quantum stress tensor.
It contains the square of concentration like the expressions for the two-particle concentration presented above.
However, the concentration square is written in formal way.
This term refer to the interaction of particles being in different quantum states
while the square of concentration includes the product of the same quantum states.
Let us introduce $\tilde{n}^{2}$,
where $\tilde{n}$ is the reduced concentration corresponding to the product of particles being in different quantum states.
Hence, equation (\ref{BECBMFMR2020 sigma via n BEC and nf}) appears as
$\sigma^{\alpha\beta}=
\frac{1}{2}g\delta^{\alpha\beta}[2\tilde{n}^{2}
+\sum_{g}n_{g}(n_{g}-1)|\varphi_{g}|^{4}]$.
The missing term (corresponding to the product of the same quantum states)
can be extracted from the second of the $\sum_{g}n_{g}^{2}|\varphi_{g}|^{4}$.
However, we need twice of it.
Therefore, the expression for the quantum stress tensor via full concentration is
$\sigma^{\alpha\beta}=
\frac{1}{2}g\delta^{\alpha\beta}[2n^{2}
-\sum_{g}n_{g}(n_{g}+1)|\varphi_{g}|^{4}]$.
We can neglect $1$ in compare with $n_{g}$ for the large occupation numbers.
The account of the product of the same quantum states does not change the expressions for $\sigma^{\alpha\beta}$.
Hence, we neglect them, but we keep the product of the same quantum state for the lowest energy state
$\sigma^{\alpha\beta}=
\frac{1}{2}g\delta^{\alpha\beta}[2n^{2}
-n_{g_{0}}^{2}|\varphi_{g}|^{4}]$,
where $g_{0}$ are the quantum numbers for the quantum state with the lowest energy.
The last term is the square of the concentration of the BEC $n_{BEC}=n_{g_{0}}|\varphi_{g}|^{2}$
in accordance with equation (\ref{BECBMFMR2020 n via varphi}).
Complete concentration $n$ is the sum of the concentration of BEC $n_{BEC}$ and the concentration of the normal fluid $n_{n}$:
$n=n_{BEC}+n_{n}$.
It gives the final expression for the quantum stress tensor for finite temperature
\begin{equation}\label{BECBMFMR2020 sigma via n BEC and nf simplified} \sigma_{BEC,n}^{\alpha\beta}=\frac{1}{2}g\delta^{\alpha\beta} \biggl(2n_{BEC}n_{n}+2n_{n}^{2}+n_{BEC}^{2}\biggr).\end{equation}

The quantum stress tensor simplifies at the zero temperature, for the BEC contribution only
\begin{equation}\label{BECBMFMR2020 sigma BEC only}
\sigma_{BEC}^{\alpha\beta}=\frac{1}{2}g\delta^{\alpha\beta}n^{2}_{BEC}.\end{equation}

It allows us present the truncated version of the Euler equation (\ref{BECBMFMR2020 Eiler eq v1}) for the BEC:
\begin{equation}\label{BECBMFMR2020 Eiler BEC}
mn(\partial_{t}+\textbf{v}\nabla)\textbf{v}
-\frac{\hbar^{2}}{2m}n\nabla\frac{\triangle\sqrt{n}}{\sqrt{n}}
=-n\nabla V_{ext}-g n\nabla n. \end{equation}

The concentration $n$ can be dropped in all terms in equation (\ref{BECBMFMR2020 Eiler BEC}).

For the eddy-free motions (the curl-free condition)
$\textbf{v}(\textbf{r},t)=\nabla\phi(\textbf{r},t)$
the second term in the Euler equation (\ref{BECBMFMR2020 Eiler BEC}) simplifies
to $(\textbf{v}\nabla)\textbf{v}=\frac{1}{2}\nabla v^{2}$.

The Euler equation (\ref{BECBMFMR2020 Eiler BEC}) is the final form of the Euler equation for the weakly-interacting BECs of neutral particles.

The GP equation can be found as the representation of hydrodynamics (\ref{BECBMFMR2020 continuity eq via v1}) and (\ref{BECBMFMR2020 Eiler BEC})
via the non-linear Schrodinger equation.

The condition of eddy-free motion allows to find
the Cauchy-Lagrangian integral of the Euler equation (\ref{BECBMFMR2020 Eiler BEC}):
\begin{equation}\label{BECBMFMR2020 Cauchy-Lagrangian integral}\partial_{t}\phi+\frac{1}{2}v^{2}+\frac{1}{m}gn -\frac{\hbar^{2}}{2m^{2}}\frac{\triangle\sqrt{n}}{\sqrt{n}}+\frac{1}{m}V_{ext}=const.
\end{equation}

The non-linear Schrodinger equation is obtained for the effective macroscopic wave function
which is constructed of the macroscopic hydrodynamic functions:
\begin{equation}\label{BECBMFMR2020 WF macroscopic}
\Phi(\textbf{r},t)=\sqrt{n(\textbf{r},t)}\exp\biggl(\frac{\imath}{\hbar}m\phi(\textbf{r},t)\biggr).
\end{equation}

We consider the time derivative of function (\ref{BECBMFMR2020 WF macroscopic}) using the hydrodynamic equations:
the continuity equation (\ref{BECBMFMR2020 continuity eq via v1}),
and
the Cauchy-Lagrangian integral of the Euler equation
(\ref{BECBMFMR2020 Cauchy-Lagrangian integral}).
After calculations we find the GP equation
\begin{equation}\label{BECBMFMR2020 GP derived}
\imath\hbar\partial_{t}\Phi=\biggl(-\frac{\hbar^{2}\nabla^{2}}{2m}+V_{ext} +g\mid\Phi\mid^{2}\biggr)\Phi.\end{equation}
It is presented in Sec. II (see eq. (\ref{BECBMFMR2020 GP})),
but we repeat it here as the result of presented derivation.

The wave function $\Phi$ is normalized on the number of particles in the system
\begin{equation}\label{BECBMFMR2020} \int d\textbf{r}\Phi(\textbf{r},t)^{*}\Phi(\textbf{r},t)=N.\end{equation}

\section{Extended set of the quantum hydrodynamic equations for the BEC: model of BEC beyond the mean-field approximation}


The method of derivation demonstrated in Sec. II can be applied to the derivation of more general models.
Let us present the most general extended QHD model obtained today for the BECs.
This model consists of four equations for tensors of different ranks:
the scalar field concentration $n$,
the vector field velocity $\textbf{v}$,
the second rank tensor field of pressure $p^{\alpha\beta}$,
and the third rank tensor field $Q^{\alpha\beta\gamma}$.

The Euler equation for the velocity field evolution is derived with the account of interaction up to the third order by the interaction radius.
While the pressure evolution equation and the third rank tensor field evolution equation $Q^{\alpha\beta\gamma}$ giving the quantum fluctuations contribution are found in the first order by the interaction radius.

The third rank tensor field $Q^{\alpha\beta\gamma}$ has the following definition in terms of microscopic parameters
\begin{equation}\label{BECBMFMR2020 def Q} Q^{\alpha\beta\gamma}(\textbf{r},t)=\int dR\sum_{i=1}^{N}\delta(\textbf{r}-\textbf{r}_{i})a^{2}(R,t)u^{\alpha}_{i}u^{\beta}_{i}u^{\gamma}_{i} .\end{equation}
It appears as the flux of the pressure (\ref{BECBMFMR2020 def pressure}) in the comoving space.

Let us present the explicit form of the four described hydrodynamic equations.

First, we repeat the continuity equation:
\begin{equation}\label{BECBMFMR2020 cont eq via v2}
\partial_{t}n+\nabla\cdot (n\textbf{v})=0. \end{equation}

The Euler equation
$$mn\partial_{t}v^{\alpha} +mn(\textbf{v}\cdot\nabla)v^{\alpha}
-\frac{\hbar^{2}}{2m}n\nabla\frac{\triangle\sqrt{n}}{\sqrt{n}}$$
$$+\partial_{\beta}p_{qf}^{\alpha\beta}
=-n\partial^{\alpha}V_{ext}
-g n\partial^{\alpha}n$$
\begin{equation}\label{BECBMFMR2020 Euler BEC 2}
-\frac{g_{2}}{2}\partial^{\alpha}\triangle n^{2}
+\frac{g_{3}}{2}  I_{2}^{\alpha\beta\gamma\delta} \partial_{\beta}\partial_{\gamma}\partial_{\delta}n^{2}\end{equation}
differs from equation (\ref{BECBMFMR2020 Eiler BEC}) by the presence of three terms.

The third and fourth term on the right-hand side are caused by the SRI included up to the TOIR approximation
\cite{Andreev PRA08}, \cite{Andreev MPL B 12}, \cite{Andreev LP 19}.
These terms contain the higher derivatives of the concentration
similarly to the quantum Bohm potential given by the third term on the left-hand side.
The isotropic part of the interaction in the TOIR approximation should be smaller then the GP term.
However, it can be comparable with the quantum Bohm potential.
Therefore, the interaction can considerably change the contribution of the "quantum" term up to changing it sign.
It can dramatically change behavior of some phenomena like the dispersive (nondissipative) shock waves \cite{Kamchatnov PRA 12}, \cite{Hoefer PRA 06}, \cite{Hoefer PRA 09}.
since the existence of higher space derivatives is crucial for their existence.

The kinetic pressure tensor $p_{qf}^{\alpha\beta}$ in the Euler equation (\ref{BECBMFMR2020 Euler BEC 2}) is kept nonzero
since it is the source of the quantum fluctuations \cite{Andreev 2005}.
The subindex $qf$ refers to the nature of this term.
Importance of the higher rank tensors for the description of classic and quantum fluid is discussed in Ref. \cite{Tokatly PRB 99}.

The Euler equation contains the following new notations:
\begin{equation} \label{BECBMFMR2020 I 4} I_{0}^{\alpha\beta\gamma\delta}=\delta^{\alpha\beta}\delta^{\gamma\delta} +\delta^{\alpha\gamma}\delta^{\beta\delta}+\delta^{\alpha\delta}\delta^{\beta\gamma}, \end{equation}
and
$I_{2}^{\alpha\beta\gamma\delta}$.
Tensor $I_{2}^{\alpha\beta\gamma\delta}$ is the nonsymmetric tensor.
It is partially symmetric.
There is the symmetry relatively permutations of the last three indexes.
However, there is no symmetry relatively permutations of the first index and other indexes.
Tensor $I_{2}^{\alpha\beta\gamma\delta}$ has the following elements:
$I_{2}^{xxxx}=I_{2}^{yyyy}=1$,
$I_{2}^{zzzz}=-2$,
$I_{2}^{xxzz}=I_{2}^{yyzz}=-2/3$,
$I_{2}^{xxyy}=I_{2}^{yyxx}=I_{2}^{zzxx}=I_{2}^{zzyy}=1/3$
and allowed permutations of indexes,
other elements are equal to zero.

Euler equation (\ref{BECBMFMR2020 Euler BEC 2}) contains the following interaction constants
\begin{equation} \label{BECBMFMR2020 def g 2} g_{2}=\frac{1}{24}\int r^{2}U(r)d\textbf{r}, \end{equation}
and
\begin{equation} \label{BECBMFMR2020 def g 3} g_{3}=\frac{1}{24\sqrt{5}}\int r^{2}U_{2}(r)d\textbf{r}. \end{equation}
Function $U_{2}$ in equation (\ref{BECBMFMR2020 def g 3}) corresponds to the anisotropic part of the SRI potential
\begin{equation}\label{BECBMFMR2020 expansion of U on Y} U(r,\theta)=\sqrt{4\pi}\sum_{l=0}^{\infty}Y_{2l,0}(\theta)U_{2l}(r),\end{equation}
where two terms give nonzero contribution in the Euler equation at the account of interaction up to the TOIR approximation:
$U(r,\theta)=U(r)+\frac{\sqrt{5}}{2}(1-3\cos^{2}\theta)U_{2}(r)$.

The quantum fluctuation reveal themselves via the nonzero value of the pressure.
The pressure evolution is calculated in the first order by the interaction radius.
It appears that there is no contribution of the interaction in this regime.
However, the evolution of the second rank tensor of pressure is caused
by the nonzero perturbations of the third rank tensor $Q_{qf}^{\alpha\beta\gamma}$:
\begin{equation} \label{BECBMFMR2020 eq evolution T qf}
\partial_{t}p_{qf}^{\alpha\beta} +\partial_{\gamma}(v^{\gamma}p_{qf}^{\alpha\beta})
+p_{qf}^{\alpha\gamma}\partial_{\gamma}v^{\beta}
+p_{qf}^{\beta\gamma}\partial_{\gamma}v^{\alpha}
+\partial_{\gamma}Q_{qf}^{\alpha\beta\gamma}=0. \end{equation}

The nonzero value of perturbations of the third rank tensor $Q_{qf}^{\alpha\beta\gamma}$ can be found
from the third rank tensor evolution equation \cite{Andreev 2005}:
$$\partial_{t}Q_{qf}^{\alpha\beta\gamma}
+\partial_{\delta}(v^{\delta}Q_{qf}^{\alpha\beta\gamma})
+Q_{qf}^{\alpha\gamma\delta}\partial_{\delta}v^{\beta}
+Q_{qf}^{\beta\gamma\delta}\partial_{\delta}v^{\alpha}$$
$$+Q_{qf}^{\alpha\beta\delta}\partial_{\delta}v^{\gamma}
-\frac{1}{mn}(p_{qf}^{\alpha\beta}\partial^{\delta}p_{qf}^{\gamma\delta}
+p_{qf}^{\alpha\gamma}\partial^{\delta}p_{qf}^{\beta\delta}$$
\begin{equation} \label{BECBMFMR2020 eq evolution Q qf}
+p_{qf}^{\beta\gamma}\partial^{\delta}p_{qf}^{\alpha\delta})
=\frac{\hbar^{2}}{4m^{3}}
g_{4}I_{0}^{\alpha\beta\gamma\delta} n\partial^{\delta}n,  \end{equation}
where
\begin{equation} \label{BECBMFMR2020 def g 4} g_{4}=\frac{2}{3}\int d\textbf{r} U''(r). \end{equation}

The right-hand side of equation (\ref{BECBMFMR2020 eq evolution Q qf}) is caused by the interaction.
It is calculated in the zero temperature regime in the first order by the interaction radius.
It characterized by the independent interaction constant $g_{4}$.

Let us consider the reduction of equations (\ref{BECBMFMR2020 eq evolution T qf}) and (\ref{BECBMFMR2020 eq evolution Q qf})
for the plane wave small perturbations $p_{qf}^{\alpha\beta}=P \exp(-\imath\omega t+\imath k x)\delta^{\alpha x}\delta^{\beta x}$
and  $Q_{qf}^{\alpha\beta\gamma}=P \exp(-\imath\omega t+\imath k x)\delta^{\alpha x}\delta^{\beta x}\delta^{\gamma x}$
with zero equilibrium values of these two parameters.

The third rank tensor evolution equation gives $Q_{qf}^{xxx}=\frac{\imath}{\omega}\frac{\hbar^{2}}{4m^{3}}
g_{4}I_{0}^{xxx\delta} n\partial^{\delta}n$,
where $I_{0}^{xxx\delta}=I_{0}^{xxxx}=3$.
We obtain $p_{qf}^{xx}=(k/\omega)Q_{qf}^{xxx}$ from the pressure evolution equation (\ref{BECBMFMR2020 eq evolution T qf}).
The Euler equation contains the following contribution of pressure $\partial_{\beta}p_{qf}^{\alpha\beta}=\imath k p_{qf}^{xx}\delta^{\alpha x}$.
Combine all of these together and find $\partial_{\beta}p_{qf}^{\alpha\beta}=-\frac{3g_{4}\hbar^{2}k^{2}}{4m^{3}\omega^{2}}n\partial^{\alpha}n$.
We also use the approximate relation between frequency $\omega$ and the wave vector $k$:
$\omega^{2}\approx \frac{gn_{0}}{m}k^{2}$.
Hence, we obtain the following reduction of the Euler equation (\ref{BECBMFMR2020 Euler BEC 2})
$$mn\partial_{t}v^{\alpha} +mn(\textbf{v}\cdot\nabla)v^{\alpha}
-\frac{\hbar^{2}}{2m}n\nabla\frac{\triangle\sqrt{n}}{\sqrt{n}}$$
\begin{equation}\label{BECBMFMR2020 Euler BEC 3} =-n\partial^{\alpha}V_{ext}
-g n\partial^{\alpha}n+\frac{3g_{4}\hbar^{2}}{4m^{2}g}\partial^{\alpha}n .\end{equation}
We drop the nonlocal terms for simplicity.

Similar extended hydrodynamic models obtained for degenerate repulsive fermions,
where the interaction is included up to the TOIR approximation,
and the set of hydrodynamic equations is extended up to the second rank tensor (the pressure evolution equation) \cite{Andreev 1912}, \cite{Andreev 2001}.


\section{Conclusion}

This mini-review has narrow area of discussion.
It is focused on the hydrodynamic description of the BECs and generalizations of the GP equation demonstrating the beyond mean-field effects.
It is focused on the models obtained within the "quantum hydrodynamics" method.
Described generalizations are based on more detailed account of interaction in the Euler equation.
However, it also includes the interaction caused effects in equations for the higher rank tensors,
where source of the quantum fluctuations is located.

Presented model demonstrates that the hydrodynamics as the method gives variety of approximations in the systematic way.
For instance, presented equations shows that
there are two sources for the additional quantum fluctuations.

If we continue our calculations in the first order by the interaction radius we can derive the equation for evolution of the fifth rank tensor
which provides nontrivial contribution proportional to the following interaction constant
$g_{5}\sim\int U^{(4)}(r)d^{3}r$,
where
$^{(4)}$ means the fourth derivative of the potential.

The second source of additional quantum fluctuations is the equation for evolution of the third rank tensor $Q^{\alpha\beta\gamma}$,
which can be generalized up to the TOIR approximation,
like it is made for the Euler equation.

Moreover, it is the open question can we obtain some contribution of interaction of bosons in the pressure evolution equation
if we consider interaction up to the TOIR at the zero temperature.

The quantum fluctuations and the terms appearing in the TOIR approximation are essential for understanding of properties of BECs
since they lead to novel phenomena (as it is described in the introduction).

Study presented in Refs. \cite{Katsimiga NJP 17 01}, \cite{Katsimiga NJP 17 02}, \cite{Katsimiga PRA 18}, \cite{Mistakidis NJP 18}
formulates the open question before the quantum hydrodynamic model.
Which approximation would be able to properly describe the soliton collisions.

\section{Acknowledgements}
Work is supported by the Russian Foundation for Basic Research (grant no. 20-02-00476).

\end{document}